# Phonon Drag Effect in Nanocomposite FeSb$_2$

Mani Pokharel[1], Huaizhou Zhao[1], Kevin Lukas[1], Bogdan Mihaila[2], Zhifeng Ren[1], and Cyril Opeil[1]

[1]Department of Physics, Boston College, Chestnut Hill, MA 02467

[2] Los Alamos National Laboratory, Los Alamos, NM 87545

**Abstract**

We study the temperature dependence of thermoelectric transport properties of four FeSb$_2$ nanocomposite samples with different grain sizes. The comparison of the single crystals and nanocomposites of varying grain size indicates the presence of substantial phonon drag effects in this system contributing to a large Seebeck coefficient at low temperature. As the grain size decreases, the increased phonon scattering at the grain boundaries leads to a suppression of the phonon-drag effect, resulting in a much smaller peak value of the Seebeck coefficient in the nanostructured bulk materials. As a consequence, the ZT values are not improved significantly even though the thermal conductivity is drastically reduced.

**Introduction**

Due to its unusual magnetic and electronic transport properties, the narrow-gap semiconductor FeSb$_2$ has been one of the extensively studied compounds in the past few decades [1-4]. The renewed interest in this compound came after Bentien *et al.* [5] reported a colossal value of the Seebeck coefficient of - 45,000 µVK$^{-1}$ with a record high value of the power factor (PF) of 2,300 µWK$^{-2}$cm$^{-1}$ at around 10 K in single crystal samples, which may make this material a potential candidate for the Peltier cooling applications at very low temperature near 10 K. Despite the huge PF value, the dimensionless figure of merit (ZT) values for single crystal samples are limited by a very high thermal conductivity $\kappa$ ~ 500 Wm$^{-1}$K$^{-1}$ at ~ 10 K. In our earlier work [6], we were able to reduce the thermal conductivity by three orders of magnitude to 0.5 Wm$^{-1}$K$^{-1}$ in achieving a peak ZT value of ~ 0.013 at ~ 50 K in nanocomposite samples. However, the Seebeck coefficient in the nanocomposites is severely degraded at low temperatures when compared to that of the single crystal counterpart. For optimized samples with $\kappa$ = 0.40 Wm$^{-1}$K$^{-1}$ and $\rho$ = 1.2 × 10$^{-4}$ $\Omega$-m at 50 K, a Seebeck coefficient of - 970 µVK$^{-1}$ is required to achieve a ZT value of 1. Unfortunately, the measured value of the Seebeck coefficient at 50 K was only - 109 µVK$^{-1}$. Therefore, it is important to know the origin of the large Seebeck coefficient in this system to further improve ZT.

The classical theory of thermoelectricity is based on the assumption that the flow of charge carriers and phonons can be treated independently. Under this assumption, the Seebeck coefficient arises due solely to spontaneous electron diffusion. However, when the two flows are linked, the effect of electron-phonon scattering should be taken into account. Hence, in general, the Seebeck coefficient is given as the sum of two independent contributions [7],

$$S = S_d + S_p,  \tag{1}$$

where $S_d$ is the conventional electron-diffusion part and $S_p$ is the phonon-drag contribution. The diffusion part is caused by the spatial variation of the electronic occupation in the presence of a thermal gradient, whereas the drag part arises due to the interaction between anisotropic lattice vibrations and mobile charge carriers. Herring [7] showed that the higher than expected value of thermopower in germanium [20] and other semiconductors could be explained assuming that the carriers are preferentially scattered by the phonons toward the cold end of the sample. The dragging of some of the charge carriers along the thermal gradient gives rise to an additional thermoelectric electromotive force (emf). Because the additional emf and the emf induced by diffusion have the same sign, the overall phonon-drag effect is to increase the Seebeck coefficient. Generally, this effect becomes stronger at lower temperatures where the phonon mean free path becomes longer.

The origin of the colossal Seebeck coefficient in $FeSb_2$ at low temperature is not completely understood yet. Several authors [5,8-11] suggested a strong electron-electron correlation as a possible cause. Other authors [12,13], however, surmised that the origin of the colossal value of the Seebeck coefficient is not due to electron-electron correlations but to the phonon drag effect. In this paper, we study the thermoelectric properties of $FeSb_2$ nanocomposites and present evidence of a substantial phonon-drag contribution to the Seebeck coefficient in $FeSb_2$.

**Experimental**

Nanostructured $FeSb_2$ samples were synthesized by first forming an ingot through melting and solidification. The ingot was ball milled and hot pressed at different temperatures to obtain the nanocomposite samples with different grain sizes ranging from nanometers to micrometers. The labels given to the different samples and the corresponding processing parameters are shown in Table 1. The Seebeck Coefficient (*S*), electrical resistivity (*ρ*), thermal conductivity (*κ*) and Hall coefficient ($R_H$) were measured on a Physical Property Measurement System (PPMS) from Quantum Design. Sample preparation and measurements were performed in the manner described in Ref. [6].

**Results and Discussions**

Figure 1 shows the temperature dependence of the Seebeck coefficient for our four nanocomposite samples. Sample S-600 was measured both in magnetic fields of 0 and 9 T. The Seebeck coefficients of all samples are significantly smaller than that of a single crystal at low temperature, whereas at high temperature the values are comparable. A room temperature value of S ~ 26 µVK$^{-1}$ was observed for all our samples. This value is comparable to S ~ 31 µVK$^{-1}$ reported for the single crystal [14]. A decrease in the Seebeck coefficient at low temperature in polycrystal [10], thin films [15] and arsenic-substituted $FeSb_2$ single crystals [8] was reported earlier. As shown in the inset of Fig. 1, the peak value of the Seebeck coefficient ($S_{max}$) decreases with decreasing grain size. This is expected for the phonon-drag effect, because the nonelectronic scattering (grain boundary scattering in the case of nanocomposites) reduces the phonon mean free path, which in turn decreases the phonon-drag contribution. A similar conclusion was drawn in Ref. [13]. Yet another example of the reduction of phonon mean free path decreasing phonon drag contribution was shown by Weber *et al*. [16]. It was demonstrated, using point contacts, that when the contact size becomes comparable to the mean free path of the relevant phonons, the phonon-drag part of the thermopower is suppressed by the boundary scattering. Here we also note that the temperature profile of the Seebeck coefficient for the samples with larger grains follows the typical behavior of phonon-drag system as suggested by Blatt [24].

The Seebeck peaks shift to the higher temperature when the grain size decreases (inset of Fig. 1). This type of size-dependent shift in the Seebeck peak is one of the characteristics of the phonon- drag dominated systems, as pointed out previously [18, 19, 20].

Usually, a small magnetothermopower is expected for the phonon-drag dominated system. For sample S-600 at 25 K, we observed a relative change in the Seebeck coefficient, $\frac{S(9T)-S(0T)}{S(0T)} = 0.059$. This small value supports the non-electronic origin of the large Seebeck coefficient in this sample.

Figure 2 shows the temperature dependence of the thermal conductivity ($\kappa$) for the four samples. Calculations based on the Widemann-Franz law show that below 50 K, more than 97% of the total thermal conductivity comes from the lattice contribution. For large-grained samples, a temperature dependence of the form $\kappa \sim T^2$ is nearly obeyed at temperatures below 50 K. As the grain size decreases, we find a gradual deviation from the $T^2$ law. Weber *et al*. [23] reported a similar $T^2$ behavior in silicon below 20 K that they attribute to strong electron-phonon scattering. For a large-grained sample, the number of available phonons interacting with the carriers is large and as the grain size decreases the number of phonons decreases causing a gradual deviation from the $T^2$ law. Evidence for electron-phonon coupling in $FeSb_2$ single crystals was also reported by Peruchhi *et al*. [21], who found a large change in phonon lifetimes using optical spectroscopy. Lazarevic *et al*. [22] later suggested that the electron-phonon coupling is temperature dependent below 40 K.

The inset of Fig. 2 shows the change in the Seebeck coefficient with thermal conductivity for 25 and 50 K. At 25 K, the Seebeck coefficient increases linearly with the thermal conductivity. The linear relationship is expected because in a phonon-drag dominated system $S_p$ is expected to follow the same temperature dependence as the phonon heat capacity so that $S_p$ and the lattice thermal conductivity ($\kappa_l$) are linearly related. This indicates that phonons play a significant role in determining the Seebeck coefficient values of our samples at around 25 K. At 50 K, however, the Seebeck coefficient decreases with increasing thermal conductivity without any obvious trend. This is an indication that the phonon-mean free path decreases and the phonon-drag effect becomes weak at higher temperature.

Significant phonon drag effects are expected to occur when the dominant phonons acquire sufficient momentum to scatter carriers across the Fermi surface [31]. In a rough approximation, $T_{max} \approx \frac{1}{10}\theta_D$, where $T_{max}$ is the temperature at which the phonon drag peak occurs and $\theta_D$ is the Debye temperature. For rutile $TiO_2$ [25], $\theta_D$ is 450-780 K and the Seebeck peak occurs at 10-30 K. For bismuth [19], another well-known phonon drag system, $\theta_D$ is 119 K, whereas the Seebeck peak occurs at 2-3 K. For $FeSb_2$ polycrystal, $\theta_D \approx$ 330-350 K was reported in Ref. [10] and 256 K in Ref. [26]. The Seebeck peak in single crystal occurs at around 10 K. For our sample S-600, the peak occurs at 25 K. Hence, comparing with the other phonon drag systems, the scaling between $T_{max}$ and $\theta_D$ in $FeSb_2$ is consistent with the phonon-drag picture.

Following Herring [7], the electron-diffusion part of the Seebeck coefficient, in µVK$^{-1}$, for a semiconductor is given by,

$$S_d = \mp 86.2 \left[ ln\frac{4.7 \times 10^{15}}{n} + \frac{3}{2} ln\frac{m^*}{m} + \frac{|\Delta E|}{kT} + \frac{3}{2} lnT \right], \quad (2)$$

where $n$ is the charge carrier density expressed in cm$^{-3}$, $m$ and $m^*$ are the bare and effective masses of the electron, respectively, and $\Delta E$ is the average energy of the transported electrons relative to the band edge. In Eq. (6), the upper and lower signs are for n-type and p-type materials, respectively.

For the case of lattice scattering by long wavelength phonons, $\frac{|\Delta E|}{kT}$ can be approximated by [7],

$$\left|\frac{\Delta E}{kT}\right| = \frac{5}{2} + r, \quad (3)$$

where the scattering parameter $r$ is taken to be -1/2. $S_d$ was calculated taking $m^* = m$ and using the charge carrier density ($n$) calculated from the Hall-coefficient measurements. The diffusion part was then subtracted from the total measured Seebeck coefficient to obtain the drag contribution shown in Fig. 2. For S-600 at 25 K, we find the diffusion value, $S_d$ = -110 µVK$^{-1}$. Using Eq. (1), $S_p$ was calculated to be -238 µVK$^{-1}$. However, for S-300 the calculated diffusion contribution turned out to be slightly larger than the measured values. So no significant phonon drag contribution is expected for this sample. This is understandable, because the phonon mean free path for sample S-300 is drastically reduced so that the dominant phonons do not carry sufficient momentum to scatter carriers. Moreover, the phonon drag is expected to weaken with increasing carrier concentration and, in fact, it has been proposed that a saturation effect occurs at high concentration level. The overestimation of the diffusion part of the Seebeck coefficient in S-300 is likely due, in part, to the approximations made in the above calculations.

The phonon-drag thermopower for semiconductors, in the first-order approximation, can be written as [7],

$$S_p = \frac{\beta v_s \lambda_p n e}{\sigma T}, \qquad (4)$$

where $v_s$ is the velocity of sound, $\lambda_p$ the mean free path of the interacting phonons, $n$ the charge carrier density, $\sigma$ the electrical conductivity, $T$ is the absolute temperature and $\beta$ is the dimensionless parameter with its value ranging from 0 to 1 depending upon the strength of the interaction. Equation (4) can be used to find approximate values of the mean free path of the phonons interacting with the electrons. The mean free path of an average phonon can also be estimated from the lattice thermal conductivity based on kinetic theory [30],

$$K_l = \frac{1}{3} c_V \widetilde{\lambda_P} v_S, \qquad (5)$$

where $K_l$ is the lattice thermal conductivity and $c_V$ the phonon contribution to the total specific heat capacity. The lattice contribution of the total thermal conductivity was calculated assuming $\kappa_{total} = \kappa_l + \kappa_e$ and $\kappa_e = L\sigma T$, where $\kappa_e$ is the electronic contribution to the total thermal conductivity. The Lorenz number in the free electron model, $L = 2.45 \times 10^{-8}$ WK$^{-2}\Omega^{-1}$ was used, $c_V$ was calculated using $C_P - \gamma T = \beta T^3$ with $\gamma = 3.98 \times 10^{-3}$ JK$^{-2}$mol$^{-1}$, as reported in Ref. [10], and the $C_P$ values were obtained from the same reference. Figure 3 shows the temperature dependence of the two length scales for samples S-600. It turns out that with $v_s$ = 3116 ms$^{-1}$ and $\beta = 0.5$, the two length scales which come from two independent calculations, are of the same order of magnitude and more importantly they behave roughly the same way as a function of temperature. This suggests that similar wavelength phonons are playing a role on both the thermal conductivity and the phonon-drag effect. A similar analysis was performed in Ref. [17] to describe the unusually large Seebeck coefficient in rutile $TiO_2$ at low temperature.

According to Keyes *et al*. [27], the value of the figure of merit ($Z_p$) that can be reached using phonon drag is rather low. Based on their argument, H. J. Goldsmid [28] shows that $Z_pT$ is less than ¼ for a bulk thermoelectric material. Ivanov *et al*. [29] reported recently a similar conclusion for low-dimensional structures. These conclusions are consistent with our earlier work [6] where, despite the dramatic reduction in the thermal conductivity by three orders of magnitude, *ZT* values were not improved significantly.

Finally, our nanocomposite data and analysis of phonon drag do not support other mechanisms that might explain the large Seebeck effects reported by Bentien *et al*. [5]. Although one cannot preclude the presence of electron-electron correlation effects, their role in this phenomenon may be a minor one. The recent analysis of electron correlations using a hybrid functional approach of Becke [32] and Hegin's *GW* functional approach [33] by Tomczak *et al*. [13] suggest that the high thermopower in $FeSb_2$ should not be understood in the context of local correlations, but rather by utilizing vertex corrections to the transport coefficients. Such vertex corrections describe the phonon-drag effect. The phonon-drag effects in $FeSb_2$ are similar to those described in p-type Ge by Geballe and Hull [20]. In a similar vein the study of magnetoresistance and Hall effect by Takahashi *et al.* [12] concludes that the large Seebeck coefficient in FeSb$_2$ is unlikely to originate from electron-electron correlations because they have an insignificant effect on the Seebeck coefficient in the low-temperature insulating regime. Our data on FeSb$_2$ nanocomposite supports their conjecture that the phonon-drag effect plays an essential role for enhancing the Seebeck coefficient in the low-temperature regime, as shown in other semiconductor materials *e.g.* InSb [34] and weakly P doped Si [23].

**Conclusions**

To summarize, in this paper we analyzed qualitatively and quantitatively several indicators of the presence of substantial electron-phonon interaction in FeSb$_2$ at low temperature. Our analysis indicates that a significant phonon-drag effect must take place in coarse-grained samples at low temperature. As the grain size decreases, the phonon drag effect becomes weaker, causing a much smaller peak value of the Seebeck coefficient in fine-grained samples. Based on these results we conclude that the phonon drag plays a significant role in the colossal value of the Seebeck coefficient inFeSb$_2$ single crystals. Therefore, the ZT values of FeSb$_2$ nanocomposite cannot be improved significantly even though the thermal conductivity can be drastically reduced.

**Acknowledgements**

The authors would like to thank J. Heremans, K. Kempa and R. Farrell, S.J. for helpful discussions and comments on the manuscript. C.O. acknowledges financial support from the Trustees of Boston College. We gratefully acknowledge funding for this work by the Department of Defense, United States Air Force Office of Scientific Research, Multi-University Research Initiative (MURI) Program under Contract # FA9550-10-1-0533.

**Table 1**. Assigned IDs, processing temperature and average grain sizes estimated from the SEM images for the four nanocomposite $FeSb_2$ samples.

| Sample code | Hot pressing temperature (°C) | Average grain size (nm) |
|---|---|---|
| S-300 | 300 | 30 |
| S-400 | 400 | 100 |
| S-500 | 500 | 350 |
| S-600 | 600 | 20,000 |

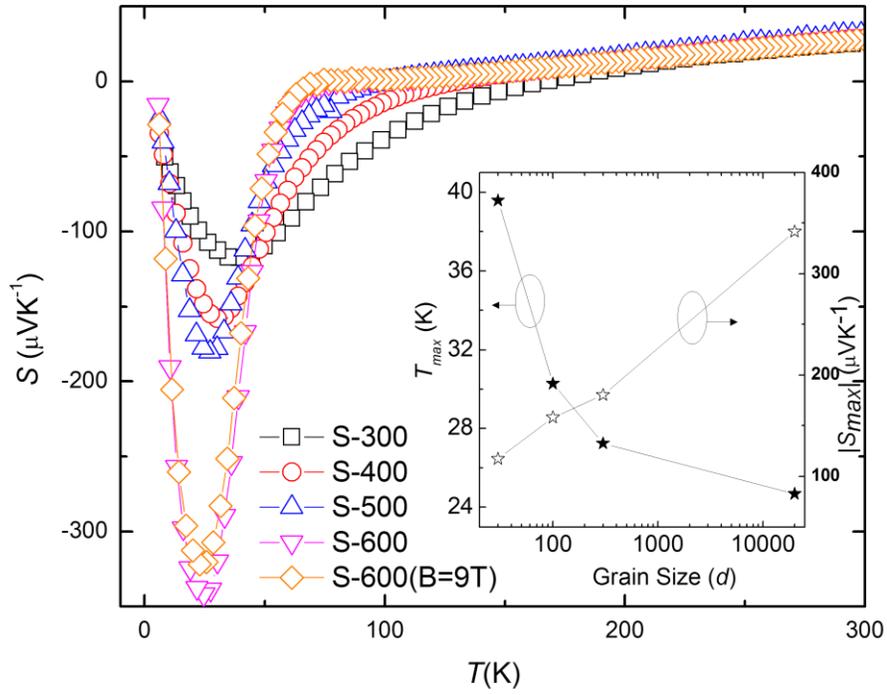

**Figure 1.** Temperature dependence of the Seebeck coefficient for the four samples. Sample S-600 was measured both at 0 and 9 Tesla magnetic fields. Inset: The grain size dependence of the peak value ($S_{max}$) and the peak position ($T_{max}$) of the Seebeck coefficient.

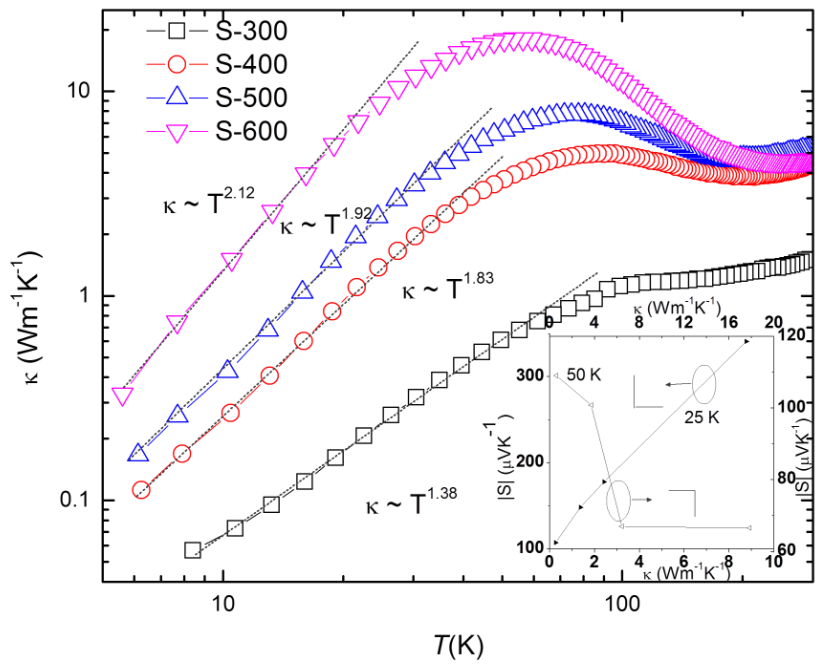

**Figure 2.** Temperature dependence of the thermal conductivity for the four samples. Fitting to the power law was done for all the samples below 50 K. Inset: Seebeck coefficient as a function of thermal conductivity at 25 and 50 K.

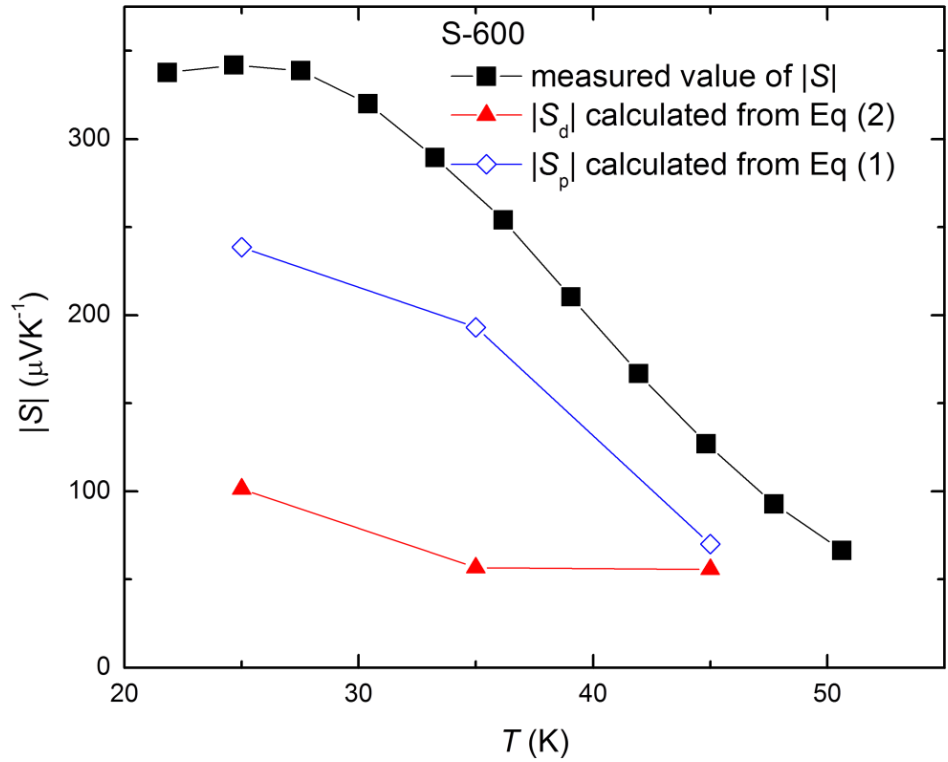

**Figure 3.** Temperature dependence of the calculated value of diffusion and drag part for the sample S-600 based on equation (1) and (2). The carrier concentration obtained from the Hall coefficient data was used in the calculation. The measured total Seebeck coefficient is also plotted for comparison.

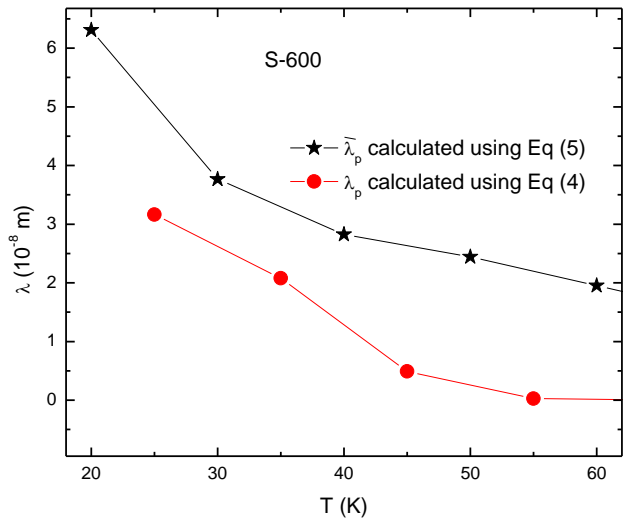

**Figure 4.** Temperature dependence of the phonon mean free paths calculated from two independent calculations using equations (4) and (5).